# Waves in the Griffiths-Podolsky metric

Kartheek R Solipuram
Chaitanya Bharathi Institute of Technology, Gandipet, Hyderabad, India. 500075

**Abstract**

Perturbations form an important section of black hole analyses. This paper deals with the effect of perturbations as in the delineation of waves that occur. It makes use of the spin coefficients from [3] to represent the general equations of waves in an accelerating black hole proposed in [2].

## I. Introduction

Since the space-time of the GP black hole is stationary and axisymmetric, one can express a general perturbation as a superposition of waves of different frequencies $\zeta$ and of different periods $(2m\pi, m = 0,1,2,.....)$ in $\varphi$. This means that one may analyze the perturbation as a superposition of different modes with a time-$t$ and a $\varphi$-dependence given by $e^{i(\zeta t + m\varphi)}$, where $m$ is an integer, positive, negative, or zero.

The basis vectors $(\vec{l}, \vec{n}, \vec{m}, \vec{\bar{m}})$ when applied as tangent vectors to the functions with a time- and a $\varphi$-dependence specified above, become the derivative operators

$$\vec{l} = D = D_0;$$
$$\vec{n} = \Delta = -\frac{\Lambda}{2\rho^2} D_0^{\neg};$$
$$\vec{m} = \delta = \frac{1}{\bar{\rho}\sqrt{2}} L_0^{\neg}; \quad (1)$$
$$\vec{\bar{m}} = \delta^* = \frac{1}{\bar{\rho}^*\sqrt{2}} L_0$$

where,

$$D_n = \partial_r + \frac{iK}{\Lambda} + \frac{2n(\partial_r \Lambda)}{\Lambda},$$
$$D_n^{\neg} = \partial_r - \frac{iK}{\Lambda} + \frac{2n(\partial_r \Lambda)}{\Lambda}, \quad (1a)$$
$$L_n = \partial_\theta + J + n\cot\theta,$$
$$L_n^{\neg} = \partial_\theta - J + n\cot\theta.$$

and

$$K = \Omega\vartheta\zeta + \Omega\varpi m,$$
$$J = (l+a)\zeta\sin\theta + m\csc\theta,$$
$$\bar{\rho} = r + i(l + a\cos\theta), \quad (1b)$$
$$\bar{\rho}^* = r - i(l + a\cos\theta),$$
$$\rho^2 = r^2 + (l + a\cos\theta)^2.$$

Here the usual convention still applies that is, while $D_n$ and $D_n^-$ are purely radial operators, $L_n$ and $L_n^-$ are purely angular operators. The differential operators that we have here satisfy a number of elementary identities which we use in the forthcoming analysis. Before proceeding further let us work out on the clarity of $\partial_r \Lambda$,

$$\partial_r \Lambda = I(Q\varpi - Qa\vartheta)\frac{\alpha l}{\omega} + \frac{(Q\varpi - Qa\vartheta)}{\Omega^2 - (P\vartheta + N\varpi)}[\Omega(UI\varpi + NIX + PI\prod + I\vartheta Y) + 2I\frac{\alpha l}{\omega}(N\varpi + P\vartheta)] + \Omega I[\wp(\varpi - a\vartheta) + Q(X - a\prod)]$$  (2)

**II. Derivation of $\partial_r \Lambda$**

Firstly, we have

$$\partial_r \Omega = -\frac{\alpha l}{\omega},$$

$$\partial_r Q = [1 + \frac{\alpha(a-l)r}{\omega}][1 - \frac{\alpha(a-l)r}{\omega}]\{(\omega^2 k + e^2 + g^2)\frac{2\alpha l}{\omega} - 2m + \frac{2\omega^2 kr}{a^2 - l^2}\} +$$

$$[(\omega^2 k + e^2 + g^2)\frac{2\alpha rl}{\omega} - 2mr + \frac{\omega^2 kr^2}{a^2 - l^2}](-\frac{2r\alpha^2(a^2 - l^2)}{\omega^2}) = \wp,$$

$$\partial_r P = \frac{(r^2 + l^2)\wp - 2r(Q - a^2)}{(r^2 + l^2)^2} = Y,$$

$$\partial_r T = \frac{(a+2l)^2}{(r^2+l^2)^2}[(r^2+l^2)\wp - 2Qr] - \frac{1}{(r^2+l^2)}[(r^2+l^2)(4r^3 + 4r(a+l)^2)] = \perp,$$

$$\partial_r N = \frac{a+2l}{(r^2+l^2)^2}[(r^2+l^2)\wp - 2Qr] + \frac{2a}{(r^2+l^2)^2}[(r^2+l^2)2r + 2(r^2 + (a+l)^2)^2] = U,$$

$$\partial_r \vartheta = \frac{(NO - PT)(YT + P\perp + a(NY + PU)) - (PT + aPN)(2U - P\perp - YT)}{(NO - PT)^2} = \prod,$$

$$\partial_r \varpi = \frac{(NO - PT)(U - aY) - (O - aP)(2U - P\perp - YT)}{(NO - PT)^2} = X,$$

$$\partial_r I = \frac{1}{\Omega^3 - \Omega(P\vartheta + N\varpi)}[\Omega(UI\varpi + NIX + PI\prod + I\vartheta Y) + 2I\frac{\alpha l}{\omega}(N\varpi + P\vartheta)],\,.$$  (3)

$$\partial_r \sqrt{Q\varpi - Qa\vartheta} = \frac{1}{\sqrt{Q\varpi - Qa\vartheta}}[\wp(\varpi - a\vartheta) + Q(X - a\prod)].$$

By definition [3],

$$\Lambda = \Omega I \sqrt{Q\varpi - Qa\vartheta} .$$ (4)

Thus,

$$\partial_r \Lambda = I(Q\varpi - Qa\vartheta)\frac{\alpha l}{\omega} + \frac{(Q\varpi - Qa\vartheta)}{\Omega^2 - (P\vartheta + N\varpi)}[\Omega(UI\varpi + NIX + PI\prod + I\vartheta Y)$$
$$+ 2I\frac{\alpha l}{\omega}(N\varpi + P\vartheta)] + \Omega I[\wp(\varpi - a\vartheta) + Q(X - a\prod)]$$ (5)

Clarified on this, we proceed to an important set of identities.

LEMMA

$$L_n(\theta) = -L_n{}^\neg(\pi - \theta),$$
$$D_n{}^\neg = (D_n)^*,$$
$$(\sin\theta)L_{n+1} = L_n \sin\theta,$$
$$(\sin\theta)L_{n+1}{}^\neg = L_n{}^\neg \sin\theta,$$  (6)
$$\Lambda D_{n+1} = D_n \Lambda,$$
$$\Lambda D_{n+1}{}^\neg = D_n{}^\neg \Lambda.$$

### III. Maxwell Equations

Maxwell's equations appropriately defined in NP formalism [1] [4] [5] are,
$$D\phi_1 - \delta^*\phi_1 = (\pi - 2\alpha)\phi_0 + 2\rho\phi_1 - \kappa\phi_2,$$
$$D\phi_2 - \delta^*\phi_1 = -\lambda\phi_0 + 2\pi\phi_1 + (\rho - 2\varepsilon)\phi_2,$$
$$\delta\phi_1 - \Delta\phi_0 = (\mu - 2\gamma)\phi_0 + 2\tau\phi_1 - \sigma\phi_2,$$  (7)
$$\delta\phi_2 - \Delta\phi_1 = -\nu\phi_0 + 2\mu\phi_1 + (\tau - 2\beta)\phi_2.$$
These in GP geometry become,

$$(D_0 - 2\rho)\phi_1 = \frac{1}{\overline{\rho}^*\sqrt{2}}(L_1 - \frac{\pi\overline{\rho}^*}{\sqrt{2}} + \frac{i\csc\theta}{2(\overline{\rho}^*)^2\overline{\rho}} - \cot\theta)\phi_0,$$

$$(D_0 - \rho + 2\varepsilon)\phi_2 = \frac{1}{\overline{\rho}^*\sqrt{2}}(L_0 + 2\sqrt{2}\pi\overline{\rho}^*)\phi_1,$$  (8)

$$(\frac{1}{\overline{\rho}\sqrt{2}}L_0{}^\neg - 2\tau)\phi_1 = (D_0{}^\neg + \mu - 2\gamma)\phi_0,$$

$$\frac{1}{\overline{\rho}\sqrt{2}}(L_1{}^\neg + \overline{\rho}\sqrt{2}(2\beta - \tau) - \overline{\rho}\sqrt{2}\cot\theta)\phi_2 = \frac{\Lambda}{2\rho^2}(-D_0 + \frac{4\rho^2}{\Lambda}\mu)\phi_1.$$

Simplifying these ones by the substitutions,

$$\aleph_0 = \phi_0,$$
$$\aleph_1 = \bar{\rho}^*\sqrt{2}\phi_1, \qquad (9)$$
$$\aleph_2 = \bar{\rho}^*\sqrt{2}\phi_2.$$

We have,

$$(D_0 - 2\rho)\aleph_1 = (L_1 - \frac{\pi\bar{\rho}^*}{\sqrt{2}} + \frac{i\csc\theta}{2(\bar{\rho}^*)^2\bar{\rho}} - \cot\theta)\aleph_0,$$

$$(D_0 - \rho + 2\varepsilon)\aleph_2 = (L_0 + 2\sqrt{2}\pi\bar{\rho}^*)\aleph_1,$$

$$(L_0^\neg - 2\bar{\rho}\sqrt{2}\tau)\aleph_1 = \Lambda(-D_1^\neg + \frac{2\rho^2}{\Lambda}(\mu - 2\gamma) + 2\frac{\partial_r\Lambda}{\Lambda})\aleph_0, \qquad (10)$$

$$(L_1^\neg + \bar{\rho}\sqrt{2}(2\beta - \tau) - \bar{\rho}\sqrt{2}\cot\theta)\aleph_2 = \Lambda(-D_0 + \frac{4\rho^2}{\Lambda}\mu)\aleph_1.$$

In this set of equations, consider the first and the third equations. They are very evidently reduced to,

$$\Lambda(D_0 - 2\rho)(D_0^\neg - \frac{2\rho^2}{\Lambda}(\mu - 2\gamma)) + (L_0^\neg - 2\bar{\rho}\sqrt{2}\tau)(L_0 - \frac{\pi\bar{\rho}^*}{\sqrt{2}} + \frac{i\csc\theta}{2(\bar{\rho}^*)^2\bar{\rho}}) = 0. \quad (11a)$$

Similarly, the second and the fourth of the equations are reduced to,

$$(L_0 + 2\sqrt{2}\pi\bar{\rho}^*)(L_1^\neg + \bar{\rho}\sqrt{2}(2\beta - \tau) - \bar{\rho}\sqrt{2}\cot\theta) + \Lambda(D_0 - \rho + 2\varepsilon)(-D_0 + \frac{4\rho^2}{\Lambda}\mu) = 0.$$

(11b)

Finally, we shall generalize these equations so that they are applicable to mass less fields of spin $|s|$,

$$\Lambda(D_{1-|s|} - 2\rho)(D_{1-|s|}^\neg - \frac{2\rho^2}{\Lambda}(\mu - 2\gamma)) + (L_{1-|s|}^\neg - 2\bar{\rho}\sqrt{2}\tau)(L_{1-|s|} - \frac{\pi\bar{\rho}^*}{\sqrt{2}} + \frac{i\csc\theta}{2(\bar{\rho}^*)^2\bar{\rho}}) = 0,$$

(12a)

$$(L_{1-|s|} + 2\sqrt{2}\pi\bar{\rho}^*)(L_1^\neg + \bar{\rho}\sqrt{2}(2\beta - \tau) - \bar{\rho}\sqrt{2}\cot\theta) + \Lambda(D_{1-|s|} - \rho + 2\varepsilon)(-D_{1-|s|} + \frac{4\rho^2}{\Lambda}\mu) = 0.$$

(12b)

Thus, these equations become the equations governing the propagation of three different kinds of waves in the GP geometry accordingly as $|s|$ is 1 for photons, $1/2$ for the two-component neutrinos, and 2 for gravitational waves.

## IV. Acknowledgements

I would like to thank KVPY [Kishore Vaigyanik Protsahan Yojna] grant provided by the Dept. of Science, Govt. of India for the culmination of this work. I would like to sincerely thank Prof. Jerry Griffiths, Loughborough University, and Prof. Naresh Dadhich, Inter University Center for Astronomy and Astrophysics [IUCAA] for their valuable mentorship in the fundamentals of this work. I also owe it to the friendly atmosphere provided to me at IUCAA in course of this work.